\documentclass[11pt]{article}

\usepackage[preprint]{acl}

\usepackage{times}
\usepackage{latexsym}

\usepackage[T1]{fontenc}
\usepackage{amsmath}
\usepackage{booktabs}
\usepackage{multirow}
\usepackage{placeins}
\usepackage[utf8]{inputenc}

\usepackage{microtype}

\usepackage{inconsolata}

\usepackage{graphicx}

\usepackage[ruled,linesnumbered]{algorithm2e}

\title{AGRO-SQL: Agentic Group-Relative Optimization \\ with High-Fidelity Data Synthesis}
\author{
    Cehua Yang\textsuperscript{1}\thanks{\ \ These authors contributed equally.},
    Dongyu Xiao\textsuperscript{1}\footnotemark[1],
    Junming Lin\textsuperscript{1},
    Yuyang Song\textsuperscript{1},
    Hanxu Yan\textsuperscript{1}, \\ 
    \textbf{Shawn Guo}\textsuperscript{2},
    \textbf{Wei Zhang}\textsuperscript{3},
    \textbf{Jian Yang}\textsuperscript{3},
    \textbf{Mingjie Tang}\textsuperscript{1}\thanks{\ \ Corresponding author.},
    \textbf{Bryan Dai}\textsuperscript{2} \\
    \textsuperscript{1}Sichuan University \quad
    \textsuperscript{2}IQuest Research \quad
    \textsuperscript{3}Beihang University \\
    \texttt{\{yangcehua, dongyuxiao, ljm12914\}@stu.scu.edu.cn} \\
    \texttt{\{yuyangsong004, hanxuyan888, tangrock\}@gmail.com} \\
    \texttt{\{zwpride, jiayang\}@buaa.edu.cn}, \texttt{\{syguo02, cbdai\}@ubiquant.com}
}

\begin{document}
\maketitle
\begin{abstract}
The advancement of Text-to-SQL systems is currently hindered by the scarcity of high-quality training data and the limited reasoning capabilities of models in complex scenarios~\cite{hu-etal-2023-importance,li2023llmservedatabaseinterface}. In this paper, we propose a holistic framework that addresses these issues through a dual-centric approach. From a \textbf{Data-Centric} perspective, we construct an iterative data factory that synthesizes RL-ready data characterized by high correctness and precise semantic-logic alignment, ensured by strict execution verification~\cite{hu-etal-2023-importance,caferoglu2025singsqlsyntheticdatageneration,dai2025reexsqlreasoningexecutionawarereinforcement}. From a \textbf{Model-Centric} perspective, we introduce a novel Agentic Reinforcement Learning framework. This framework employs a \textbf{Diversity-Aware Cold Start} stage to initialize a robust policy, followed by \textbf{Group Relative Policy Optimization (GRPO)} to refine the agent's reasoning via environmental feedback~\cite{deepseekmath2024,zhang2025edgegrpoentropydrivengrpoguided}. Extensive experiments on BIRD~\cite{li2023llmservedatabaseinterface} and Spider~\cite{yu2018spider} benchmarks demonstrate that our synergistic approach achieves state-of-the-art performance among single-model methods.
\end{abstract}

\section{Introduction}
\label{sec:intro}

Text-to-SQL aims to democratize database access by translating natural language into executable queries~\cite{pourreza-rafiei-2023-dinsql,li2023llmservedatabaseinterface}. While Large Language Models (LLMs) have demonstrated impressive capabilities, training specialized, efficient models faces significant bottlenecks on realistic benchmarks such as BIRD~\cite{li2023llmservedatabaseinterface} and Spider~\cite{yu2018spider}. We identify two primary hurdles: (1) \textbf{Data Scarcity and Quality}: High-quality, complex Text-to-SQL pairs are expensive to annotate, and existing datasets often lack the scale and precision required for stable Reinforcement Learning (RL)~\cite{hu-etal-2023-importance,li2023llmservedatabaseinterface}. (2) \textbf{Reasoning Limitations}: Standard Supervised Fine-Tuning (SFT) often fails to imbue models with self-correction capabilities, while standard RL methods frequently struggle with instability and inefficient exploration in the sparse-reward environment of SQL generation~\cite{sheng2025cscsqlcorrectiveselfconsistencytexttosql,pourreza2025reasoningsqlreinforcementlearningsql}.

In this work, we present a unified framework that tackles these challenges simultaneously through \textbf{Data-Centric} and \textbf{Model-Centric} pathways.

\textbf{Data-Centric: RL-Ready Synthesis.} To break the data scaling wall, we design an iterative data factory to synthesize interactive trajectories~\cite{hu-etal-2023-importance,caferoglu2025singsqlsyntheticdatageneration}. To ensure the generated samples are suitable for RL, we enforce strict \textbf{Semantic-Logic Alignment}: we apply a ``Generation-as-Verification'' strategy, retaining only those trajectories where the execution results perfectly match the ground truth~\cite{dai2025reexsqlreasoningexecutionawarereinforcement,weng2025graphrewardsqlexecutionfreereinforcementlearning}. This yields a massive scale of high-correctness data, preventing reward hacking during the subsequent RL stage~\cite{dai2025reexsqlreasoningexecutionawarereinforcement}.

\textbf{Model-Centric: Agentic RL with GRPO.} The availability of complex synthetic data exposes the limitations of traditional training methods~\cite{pourreza-rafiei-2023-dinsql,chase_sql_2024}. To bridge the gap, we propose a two-stage Agentic RL framework. First, we implement a \textbf{Diversity-Aware Cold Start} to initialize a robust policy from curated high-quality trajectories~\cite{hu-etal-2023-importance,caferoglu2025singsqlsyntheticdatageneration}. Second, we optimize the agent using \textbf{Group Relative Policy Optimization (GRPO)}~\cite{deepseekmath2024,zhang2025edgegrpoentropydrivengrpoguided}. Unlike standard methods that rely on unstable value networks, GRPO iteratively refines the policy by comparing the relative execution rewards of a group of synthesized trajectories, stabilizing training and improving exploration under sparse feedback~\cite{deepseekmath2024,cheng2025reasoningexplorationentropyperspective}.

Our contributions are summarized as follows:
\begin{itemize} 
\item We build a data pipeline that alleviates data scarcity by synthesizing high-correctness, RL-ready data via strict execution verification.
\item We propose an Agentic RL framework combining Diversity-Aware Cold Start and GRPO to enhance reasoning and exploration~\cite{deepseekmath2024,zhang2025edgegrpoentropydrivengrpoguided}.
\item We demonstrate that our method achieves single-model SOTA results on BIRD~\cite{li2023llmservedatabaseinterface} and Spider~\cite{yu2018spider} benchmarks.
\end{itemize}

\section{Related Work}

\subsection{Data Synthesis in Text-to-SQL}

Data synthesis alleviates annotation bottlenecks in Text-to-SQL by generating additional NL-SQL pairs via templates, schema-guided sampling, and LLM generation. Large-scale synthetic corpora enable pretraining and SFT, improving coverage across domains and dialects~\cite{hu-etal-2023-importance, li-etal-2025-omnisql, pourreza2024sqlgenbridgingdialectgap}. Because synthetic pairs can be illogical or misaligned, recent pipelines emphasize data verification, such as executability checks, relationship preservation, and automatic repair, before mixing synthetic data with human data~\cite{hu-etal-2023-importance, caferoglu2025singsqlsyntheticdatageneration}.

\subsection{Reinforcement Learning in Text-to-SQL}
Supervised fine-tuning for Text-to-SQL suffers from a mismatch between loss and evaluation: cross-entropy optimizes string-level overlap, while benchmarks measure execution accuracy~\cite{zhong2018seqsql}. These limitations motivate the use of reinforcement learning. Early works~\cite{zhong2018seqsql} use execution feedback as a reward to align training with correctness. However, the binary execution reward is inherently sparse and provides little learning signal for near-correct outputs. Recent works~\cite{sheng2025cscsqlcorrectiveselfconsistencytexttosql,pourreza2025reasoningsqlreinforcementlearningsql} address this by reward shaping, designing partial reward components to densify feedback. Yet even with shaped rewards, exploration is still constrained in the large SQL structural space.

\subsection{Inference-Time Strategies for Text-to-SQL}

Inference for Text-to-SQL increasingly leans on constructing richer context and reducing noise. Practical systems augment schemas with field metadata/descriptions and representative values to reduce ambiguity~\cite{shkapenyuk2025automaticmetadataextractiontexttosql, chess_2024}, as well as pruning schemas with retrieval-based schema linking~\cite{liu2025xiyansqlnovelmultigeneratorframework, chase_sql_2024, pourreza-rafiei-2023-dinsql}. Recent pipelines also adopt multi-turn refinement uses execution feedback to iteratively refine outputs~\cite{xu2025mtirsqlmultiturntoolintegratedreasoning}.

\section{Methodology}
\label{sec:method}

To address the scaling bottlenecks and logical inconsistencies discussed in Section~\ref{sec:intro}, we propose \textbf{AGRO-SQL}, an end-to-end framework that synchronizes a high-fidelity data factory with an entropy-guided optimization strategy.

\begin{figure*}[htbp]
    \centering
    \includegraphics[width=\textwidth]{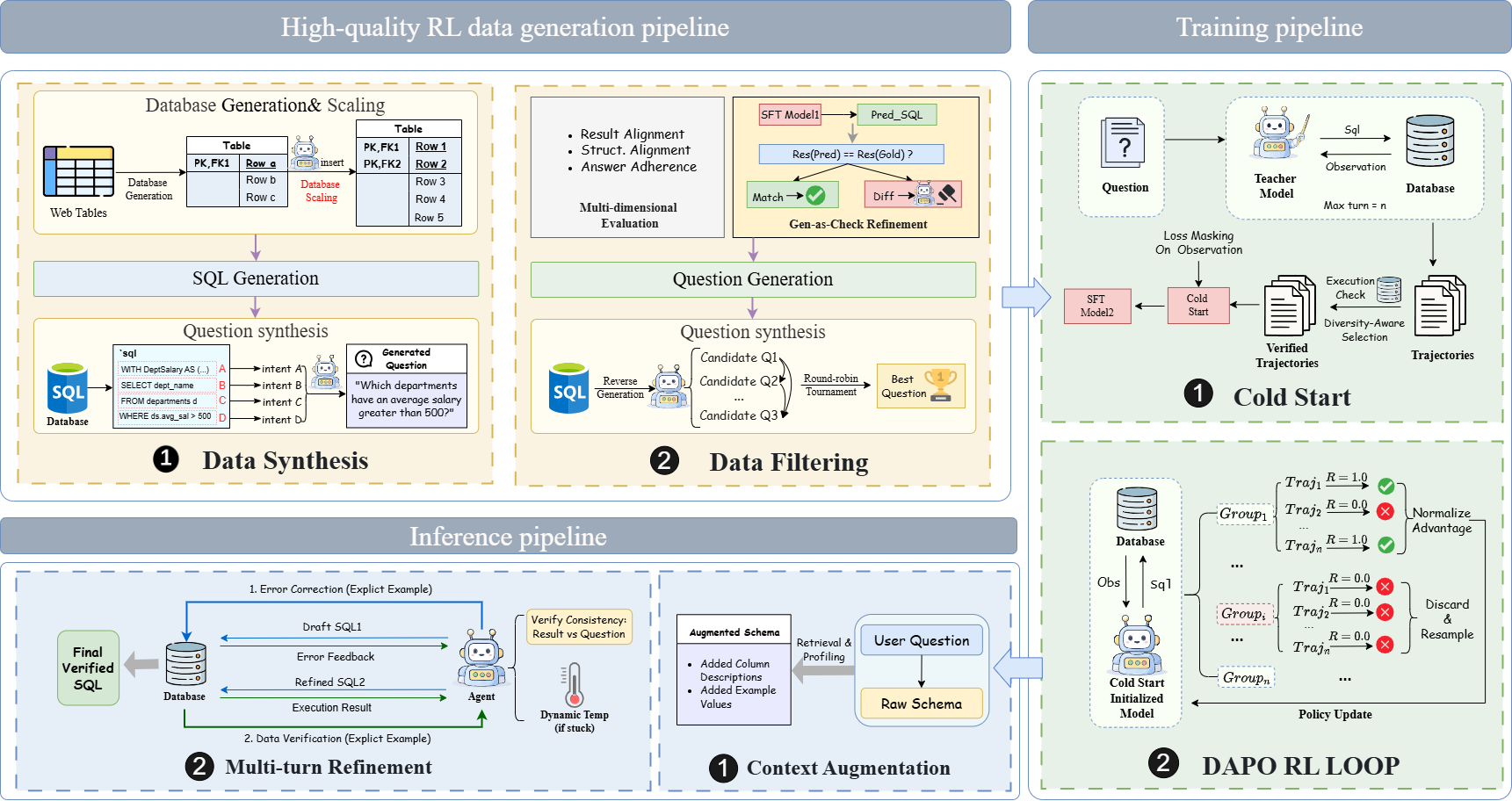}
    \caption{The overall pipeline of our framework. Given a natural language question and database schema, the policy model generates SQL candidates. Our core Advantage Shaping Module then computes reshaped token-level advantages, which are used to update the policy model via the GRPO algorithm.}
    \label{fig:pipeline}
\end{figure*}

\subsection{RL-Ready Iterative Data Pipeline}
\label{sec:data_pipeline}

To address the scaling bottlenecks and logical inconsistencies discussed in Section~\ref{sec:intro}, we propose \textbf{AGRO-SQL}, an end-to-end framework synchronizing a high-fidelity data factory with an entropy-guided optimization strategy (see Figure~\ref{fig:pipeline}). To ensure the ``zero-noise'' gold labels essential for scaling RL, we develop an \textbf{RL-Ready Iterative Data Pipeline} comprising two stages. First, the \textbf{Structural-Aware Synthesis} stage enhances baseline synthesis~\citep{li-etal-2025-omnisql} by employing DAG-based database augmentation to mitigate accidental execution correctness, enforcing SQL decomposition to capture structural constraints, and utilizing tournament-based selection to ensure high logical fidelity. Second, to eliminate ``logic noise,'' we implement a $K$-cycle \textbf{Iterative Gen-as-Check Refinement} loop. Synthesized samples are verified by comparing SFT model predictions against gold SQLs; divergences trigger a multi-dimensional audit by an \textbf{LLM-as-a-Judge}, and problematic samples are iteratively regenerated until logical consistency is confirmed via execution.

\subsection{Agentic RL Training Framework}
To enable robust multi-turn reasoning, we propose a two-stage framework comprising \textbf{Cold Start Supervised Fine-Tuning (SFT)} and \textbf{Agentic Reinforcement Learning (Agentic RL)}. In the SFT stage, we distill interactive capabilities from a teacher model, DeepSeek V3.2\cite{deepseekv3_2_2025} by synthesizing diverse trajectories. To mitigate overfitting, we employ a \textbf{diversity-aware selection mechanism} based on hybrid embeddings of SQL actions and reasoning thoughts, fine-tuning the model with a loss-masking objective that focuses solely on agent-generated tokens. In the Agentic RL stage, we further optimize the policy $\pi_\theta$ using Group Relative Policy Optimization algorithm (GRPO)\cite{deepseekmath2024}. Formulated as a POMDP, the training aligns the agent with environmental feedback using a sparse reward function ($R=1.0$ for correct execution, $R=-1.0$ for invalid format). GRPO stabilizes optimization by computing advantages $\hat{A}_i$ relative to a group of sampled trajectories, effectively encouraging self-correction and semantic accuracy.

\subsection{Inference}

To mitigate semantic-logic mismatch, we propose a three-stage context augmentation pipeline. \textbf{Database Profile Construction:} The pipeline starts by profiling each column's statistical exemplars~\cite{shkapenyuk2025automaticmetadataextractiontexttosql, chess_2024}. \textbf{Semantic Description Generation:} We first harvest any existing metadata from the benchmark or database. If not, we generate descriptions using LLM. \textbf{Dynamic Pruning via Retrieval:} Then we conduct context pruning, using a bi-encoder to calculate embeddings and retrieve top-K units~\cite{reimers-gurevych-2019-sentence}. Finally, to keep critical columns for potential multi-hop joins, we remove them from the retrieval process and always retain them~\cite{wang-etal-2020-rat, wang-etal-2025-linkalign}.

We implement a Multi-turn Refinement mechanism with Execution Feedback to transform the task into a action-feedback loop for refining SQL. At each turn, the LLM agent observes the history and generates an action. The environment executes the SQL and returns an observation. This cycle enables the model to correct syntactic and schema Errors, as well as semantic divergences.

\begin{table*}[!t]
    \centering
    \caption{Performance comparison on two Text-to-SQL benchmarks. We report Execution Accuracy (EX, \%) for BIRD and Spider. \textbf{Bold} denotes our method}
    \label{tab:main_results}
    \resizebox{\textwidth}{!}{
        \begin{tabular}{l c c c c c}
            \toprule
            \textbf{Model} & \textbf{Base Model} & \textbf{Training Set} & \textbf{Strategy} & \textbf{Bird} & \textbf{Spider} \\
            \midrule
            GPT-5.1~\cite{openai_gpt5} & -- & -- & Zero-shot & 53.31 & 77.60 \\
            O3-mini~\cite{openai_o3mini} & -- & -- & Zero-shot & 61.34 & 78.82 \\
            Claude-Opus-4-5-20251101~\cite{anthropic_claude_opus45_20251101} & -- & -- & Zero-shot & 66.01 & 76.0 \\ 
            Gemini-3-Flash-preview~\cite{gcp_vertex_gemini3_flash} & -- & -- & Zero-shot & 66.60 & 87.20 \\ 
            Gemini-3-Pro-preview~\cite{gcp_vertex_gemini3_pro} & -- & -- & Zero-shot & 67.52 & 87.00 \\ 
            Qwen3-8B-Base~\cite{qwen3_report} & Qwen3-8B-Base & -- & Zero-shot & 45.15 & 68.38 \\
            Qwen2.5-Coder-32B~\cite{qwen25_coder_report} & Qwen2.5-Coder-32B & -- & Instruction & 62.10 & 83.90 \\
            
            \midrule
            \multicolumn{6}{c}{\textit{Multi-agent Framework}} \\
            \midrule
            CHESS$^{\dagger}$~\cite{chess_2024} & Gemini-1.5-pro & -- & Multi-agent & 68.31 & 87.20 \\
            CHASE-SQL$^{\dagger}$~\cite{chase_sql_2024} & Gemini-1.5-pro & -- & Multi-agent & 73.01 & 87.60 \\
            OpenSearch-SQL$^{\dagger}$~\cite{10.1145/3725331} &  GPT-4o-0513 & -- & Multi-agent & 69.30 & 87.10 \\
            XiYan-SQL$^{\dagger}$~\cite{liu2025xiyansqlnovelmultigeneratorframework} & Qwen-2.5-Coder-32B & B+S$^{*}$ & Multi-agent & 73.34 & 89.65 \\
            \midrule
            \multicolumn{6}{c}{\textit{Single SFT \& Agentic Model }} \\
            \midrule
            Arctic-T2S-R1~\cite{yao2025arctictext2sqlr1simplerewardsstrong} & Qwen2.5-Coder-7B-Instruct & B+S$^{*}$ & RL & 58.93 & 80.75 \\
            OmniSQL~\cite{li-etal-2025-omnisql} & Qwen2.5-Coder-32B-Instruct & OmniSQL & SFT & 58.80 & 79.01 \\
            Reasoning-SQL$^{\dagger}$~\cite{pourreza2025reasoningsqlreinforcementlearningsql} & Qwen2.5-Coder-14B-Instruct & BIRD & RL & 65.31 & 81.43\\
            MARS-SQL~\cite{mars_sql_2025} & Qwen2.5-Coder-7B-Instruct & BIRD & Agentic RL & 57.32 & 79.04 \\
            \midrule
            Qwen3-8B-Base + SFT & Qwen3-8B-Base & BIRD & SFT & 62.65 & 88.39 \\
            Qwen3-8B-Base + SFT + RL & Qwen3-8B-Base & BIRD & RL & 63.17 & 89.26 \\
            \textbf{AGRO-SQL (Ours)} & Qwen3-8B-Base & BIRD & Agentic RL & \textbf{72.10} & \textbf{89.13} \\
            \bottomrule
            \multicolumn{6}{l}{\footnotesize $^{\dagger}$ Results are cited from original papers as their source code or specific inference frameworks are not publicly available.} \\
            \multicolumn{6}{l}{\footnotesize $^{*}$ Trained on the union of the BIRD and SPIDER training sets.}

        \end{tabular}
    }
    
\end{table*}

\section{Experiments}
In this section, we evaluate the effectiveness of our proposed method. We first introduce the experimental setup, followed by a presentation and analysis of the main results. 

\subsection{Datasets}
We conduct experiments on two widely-recognized Text-to-SQL benchmarks to ensure a comprehensive evaluation of our model's capabilities.

\textbf{Spider}~\citep{yu2018spider} is a foundational and widely-used benchmark in the Text-to-SQL field. It contains 7,000 training samples and 1,034 development samples, spanning 200 databases across 138 diverse domains.

\textbf{Bird}~\citep{li2023llmservedatabaseinterface} represents a more recent and challenging benchmark, designed to better reflect real-world application scenarios. This dataset features a larger scale, with 9,428 training and 1,534 development samples. It covers 95 large-scale databases from 37 professional domains, demanding more complex reasoning than Spider.
\subsection{Metrics}
We use Execution Accuracy (EX)  to evaluate our model. EX serves to estimate the proportion of questions that produce identical executed results for predicted and ground-truth SQLs.

\subsection{Baselines}
We compare AGRO-SQL with a broad range of baselines organized into three groups:
\begin{itemize}
    \item \textbf{Base Models:} We evaluate several strong LLMs under zero-shot prompting,
    including O3-mini~\cite{openai_o3mini}, GPT-5.1~\cite{openai_gpt5},
    Claude-Opus-4-5-20251101~\cite{anthropic_claude_opus45_20251101},
    Gemini-3-Flash-preview~\cite{gcp_vertex_gemini3_flash} and Gemini-3-Pro-preview~\cite{gcp_vertex_gemini3_pro},
    as well as open-weight baselines Qwen3-Base-8B~\cite{qwen3_report} and Qwen2.5-Coder-32B~\cite{qwen25_coder_report}.

    \item \textbf{Closed-Source Frameworks:} We include representative multi-agent Text-to-SQL systems
    that leverage proprietary LLMs for query generation, including CHESS~\cite{chess_2024},
    CHASE-SQL~\cite{chase_sql_2024}, OpenSearch-SQL~\cite{10.1145/3725331},
    and XiYan-SQL~\cite{liu2025xiyansqlnovelmultigeneratorframework}.

    \item \textbf{Open-Source Agentic Model \& Framework:} We compare against open-source and post-trained Text-to-SQL models,
    including our Qwen3-Base-8B SFT baseline and its RL variant (trained on BIRD),
    as well as Arctic-Text2SQL-R1~\cite{yao2025arctictext2sqlr1simplerewardsstrong},
    OmniSQL~\cite{li-etal-2025-omnisql},
    Reasoning-SQL~\cite{pourreza2025reasoningsqlreinforcementlearningsql},
    and MARS-SQL~\cite{mars_sql_2025}.
\end{itemize}

\subsection{Implementation Details}
We implement our method using Qwen3-8B-Base~\cite{qwen3_report} as our backbone model. All models were trained on a single node of NVIDIA H800 GPUs. For SFT training, we adopted a two-stage strategy: first adapting the model to the Text-to-SQL task using our synthesized large-scale dataset\cite{li-etal-2025-omnisql}, followed by further fine-tuning to align with the agentic workflow using agent interaction trajectories. For vanilla RL, the model was trained using either the BIRD dataset~\cite{li2023llmservedatabaseinterface} or our synthesized high-quality text-to-SQL dataset with GRPO~\cite{deepseekmath2024,zhang2025edgegrpoentropydrivengrpoguided}. For agentic RL, the model was trained on BIRD~\cite{li2023llmservedatabaseinterface} with GRPO~\cite{deepseekmath2024} and DAPO, with generation temperature of 0.7, a learning rate of $5\times10^{-6}$, and a total batch size of 256 (10 rollouts each).

\subsection{Main Results}
Table~\ref{tab:main_results} presents the primary results of our method against SOTA baselines on the development sets of Spider~\cite{yu2018spider} and BIRD~\cite{li2023llmservedatabaseinterface}. AGRO-SQL achieves the best performance among all single models of similar size. Notably, on the more challenging BIRD benchmark~\cite{li2023llmservedatabaseinterface}, our method shows a significant improvement in Execution Accuracy (EX), highlighting the effectiveness of the agentic RL framework in complex multi-step reasoning.

\section{Conclusion}
We presented \textbf{AGRO-SQL}, a framework synergizing high-fidelity data synthesis with agentic reinforcement learning (GRPO). By enforcing strict execution verification in data generation and stabilizing exploration via group-relative optimization, our method significantly improves reasoning robustness. Experiments on the BIRD benchmark show our agentic model achieves an execution accuracy of \textbf{70.66\%}, which further improves to \textbf{72.10\%} with self-consistency, establishing a new state-of-the-art for open-source models.

\section{Limitations}
Our approach relies heavily on executable environments for data verification, reward computation, and agentic refinement. While execution-based signals provide reliable supervision, they require access to runnable databases and introduce additional computational overhead, which may limit applicability in restricted or latency-sensitive settings. In addition, although our iterative synthesis pipeline enforces strict execution correctness, synthetic data may still exhibit coverage gaps for rare SQL patterns or long-tail schemas, and execution equivalence alone cannot fully guarantee natural language faithfulness. Addressing these limitations will require more execution-free or structure-aware feedback signals and broader validation in real-world deployment scenarios.

\section{Ethical Considerations}
Deploying autonomous Text-to-SQL agents requires strict security measures. To prevent unauthorized data exposure or modification, such systems must operate with read-only permissions and robust access controls. We also emphasize the importance of monitoring synthetic training data to mitigate potential biases in generated queries.

\bibliography{custom}

@misc{zhong2018seqsql,
    title={Seq2{SQL}: Generating Structured Queries From Natural Language Using Reinforcement Learning },
    author={Victor Zhong and Caiming Xiong and Richard Socher},
    url={https://openreview.net/forum?id=Syx6bz-Ab},
    year={2018},
}

@misc{sheng2025cscsqlcorrectiveselfconsistencytexttosql,
      title={CSC-SQL: Corrective Self-Consistency in Text-to-SQL via Reinforcement Learning}, 
      author={Lei Sheng and Shuai-Shuai Xu},
      year={2025},
      eprint={2505.13271},
      archivePrefix={arXiv},
      primaryClass={cs.CL},
      url={https://arxiv.org/abs/2505.13271}, 
}

@misc{zhang2025edgegrpoentropydrivengrpoguided,
      title={EDGE-GRPO: Entropy-Driven GRPO with Guided Error Correction for Advantage Diversity}, 
      author={Xingjian Zhang and Siwei Wen and Wenjun Wu and Lei Huang},
      year={2025},
      eprint={2507.21848},
      archivePrefix={arXiv},
      primaryClass={cs.AI},
      url={https://arxiv.org/abs/2507.21848}, 
}

@misc{cheng2025reasoningexplorationentropyperspective,
      title={Reasoning with Exploration: An Entropy Perspective on Reinforcement Learning for LLMs}, 
      author={Daixuan Cheng and Shaohan Huang and Xuekai Zhu and Bo Dai and Wayne Xin Zhao and Zhenliang Zhang and Furu Wei},
      year={2025},
      eprint={2506.14758},
      archivePrefix={arXiv},
      primaryClass={cs.CL},
      url={https://arxiv.org/abs/2506.14758}, 
}

@misc{dai2025reexsqlreasoningexecutionawarereinforcement,
      title={ReEx-SQL: Reasoning with Execution-Aware Reinforcement Learning for Text-to-SQL}, 
      author={Yaxun Dai and Wenxuan Xie and Xialie Zhuang and Tianyu Yang and Yiying Yang and Haiqin Yang and Yuhang Zhao and Pingfu Chao and Wenhao Jiang},
      year={2025},
      eprint={2505.12768},
      archivePrefix={arXiv},
      primaryClass={cs.CL},
      url={https://arxiv.org/abs/2505.12768}, 
}

@misc{weng2025graphrewardsqlexecutionfreereinforcementlearning,
      title={Graph-Reward-SQL: Execution-Free Reinforcement Learning for Text-to-SQL via Graph Matching and Stepwise Reward}, 
      author={Han Weng and Puzhen Wu and Cui Longjie and Yi Zhan and Boyi Liu and Yuanfeng Song and Dun Zeng and Yingxiang Yang and Qianru Zhang and Dong Huang and Xiaoming Yin and Yang Sun and Xing Chen},
      year={2025},
      eprint={2505.12380},
      archivePrefix={arXiv},
      primaryClass={cs.LG},
      url={https://arxiv.org/abs/2505.12380}, 
}

@misc{pourreza2025reasoningsqlreinforcementlearningsql,
      title={Reasoning-SQL: Reinforcement Learning with SQL Tailored Partial Rewards for Reasoning-Enhanced Text-to-SQL}, 
      author={Mohammadreza Pourreza and Shayan Talaei and Ruoxi Sun and Xingchen Wan and Hailong Li and Azalia Mirhoseini and Amin Saberi and Sercan "O. Arik},
      year={2025},
      eprint={2503.23157},
      archivePrefix={arXiv},
      primaryClass={cs.LG},
      url={https://arxiv.org/abs/2503.23157}, 
}

@inproceedings{hu-etal-2023-importance,
    title = "Importance of Synthesizing High-quality Data for Text-to-{SQL} Parsing",
    author = "Hu, Yiqun  and
      Zhao, Yiyun  and
      Jiang, Jiarong  and
      Lan, Wuwei  and
      Zhu, Henghui  and
      Chauhan, Anuj  and
      Li, Alexander Hanbo  and
      Pan, Lin  and
      Wang, Jun  and
      Hang, Chung-Wei  and
      Zhang, Sheng  and
      Guo, Jiang  and
      Dong, Mingwen  and
      Lilien, Joseph  and
      Ng, Patrick  and
      Wang, Zhiguo  and
      Castelli, Vittorio  and
      Xiang, Bing",
    editor = "Rogers, Anna  and
      Boyd-Graber, Jordan  and
      Okazaki, Naoaki",
    booktitle = "Findings of the Association for Computational Linguistics: ACL 2023",
    month = jul,
    year = "2023",
    address = "Toronto, Canada",
    publisher = "Association for Computational Linguistics",
    url = "https://aclanthology.org/2023.findings-acl.86/",
    doi = "10.18653/v1/2023.findings-acl.86",
    pages = "1327--1343",
}

@article{li-etal-2025-omnisql,
    author = {Li, Haoyang and Wu, Shang and Zhang, Xiaokang and Huang, Xinmei and Zhang, Jing and Jiang, Fuxin and Wang, Shuai and Zhang, Tieying and Chen, Jianjun and Shi, Rui and Chen, Hong and Li, Cuiping},
    title = {OmniSQL: Synthesizing High-Quality Text-to-SQL Data at Scale},
    year = {2025},
    issue_date = {July 2025},
    publisher = {VLDB Endowment},
    volume = {18},
    number = {11},
    issn = {2150-8097},
    url = {https://doi.org/10.14778/3749646.3749723},
    doi = {10.14778/3749646.3749723},
    journal = {Proc. VLDB Endow.},
    month = jul,
    pages = {4695–4709},
    numpages = {15},
    doi = {10.14778/3749646.3749723},
}

@misc{pourreza2024sqlgenbridgingdialectgap,
    title={SQL-GEN: Bridging the Dialect Gap for Text-to-SQL Via Synthetic Data And Model Merging}, 
    author={Mohammadreza Pourreza and Ruoxi Sun and Hailong Li and Lesly Miculicich and Tomas Pfister and Sercan O. Arik},
    year={2024},
    eprint={2408.12733},
    archivePrefix={arXiv},
    primaryClass={cs.AI},
    url={https://arxiv.org/abs/2408.12733}, 
}

@misc{caferoglu2025singsqlsyntheticdatageneration,
    title={SING-SQL: A Synthetic Data Generation Framework for In-Domain Text-to-SQL Translation},
    author={Hasan Alp Cafero{\u{g}}lu and Mehmet Serhat {\c{C}}elik and {\"{O}}zg{\"{u}}r Ulusoy},
    year={2025},
    eprint={2509.25672},
    archivePrefix={arXiv},
    primaryClass={cs.AI},
    url={https://arxiv.org/abs/2509.25672},
}

@inproceedings{pourreza-rafiei-2023-dinsql,
    title = {DIN-SQL: Decomposed In-Context Learning of Text-to-SQL with Self-Correction},
    author = {Pourreza, Mohammadreza and Rafiei, Davood},
    booktitle = {Advances in Neural Information Processing Systems},
    editor = {A. Oh and T. Naumann and A. Globerson and K. Saenko and M. Hardt and S. Levine},
    pages = {36339--36348},
    publisher = {Curran Associates, Inc.},
    url = {https://proceedings.neurips.cc/paper_files/paper/2023/file/72223cc66f63ca1aa59edaec1b3670e6-Paper-Conference.pdf},
    volume = {36},
    year = {2023}
}

@misc{liu2025xiyansqlnovelmultigeneratorframework,
      title={XiYan-SQL: A Novel Multi-Generator Framework For Text-to-SQL}, 
      author={Yifu Liu and Yin Zhu and Yingqi Gao and Zhiling Luo and Xiaoxia Li and Xiaorong Shi and Yuntao Hong and Jinyang Gao and Yu Li and Bolin Ding and Jingren Zhou},
      year={2025},
      eprint={2507.04701},
      archivePrefix={arXiv},
      primaryClass={cs.CL},
      url={https://arxiv.org/abs/2507.04701}, 
}

@misc{shkapenyuk2025automaticmetadataextractiontexttosql,
      title={Automatic Metadata Extraction for Text-to-SQL}, 
      author={Vladislav Shkapenyuk and Divesh Srivastava and Theodore Johnson and Parisa Ghane},
      year={2025},
      eprint={2505.19988},
      archivePrefix={arXiv},
      primaryClass={cs.DB},
      url={https://arxiv.org/abs/2505.19988}, 
}

@inproceedings{li2023llmservedatabaseinterface,
    author = {Li, Jinyang and Hui, Binyuan and Qu, Ge and Yang, Jiaxi and Li, Binhua and Li, Bowen and Wang, Bailin and Qin, Bowen and Geng, Ruiying and Huo, Nan and Zhou, Xuanhe and Chenhao, Ma and Li, Guoliang and Chang, Kevin and Huang, Fei and Cheng, Reynold and Li, Yongbin},
    booktitle = {Advances in Neural Information Processing Systems},
    editor = {A. Oh and T. Naumann and A. Globerson and K. Saenko and M. Hardt and S. Levine},
    pages = {42330--42357},
    publisher = {Curran Associates, Inc.},
    title = {Can LLM Already Serve as A Database Interface? A BIg Bench for Large-Scale Database Grounded Text-to-SQLs},
    url = {https://proceedings.neurips.cc/paper_files/paper/2023/file/83fc8fab1710363050bbd1d4b8cc0021-Paper-Datasets_and_Benchmarks.pdf},
    volume = {36},
    year = {2023}
}

@article{yu2018spider,
  title={Spider: A large-scale human-labeled dataset for complex and cross-domain semantic parsing and text-to-sql task},
  author={Yu, Tao and Zhang, Rui and Yang, Kai and Yasunaga, Michihiro and Wang, Dongxu and Li, Zifan and Ma, James and Li, Irene and Yao, Qingning and Roman, Shanelle and others},
  journal={arXiv preprint arXiv:1809.08887},
  year={2018}
}

@misc{xu2025mtirsqlmultiturntoolintegratedreasoning,
      title={MTIR-SQL: Multi-turn Tool-Integrated Reasoning Reinforcement Learning for Text-to-SQL}, 
      author={Zekun Xu and Siyu Xia and Chuhuai Yue and Jiajun Chai and Mingxue Tian and Xiaohan Wang and Wei Lin and Haoxuan Li and Guojun Yin},
      year={2025},
      eprint={2510.25510},
      archivePrefix={arXiv},
      primaryClass={cs.AI},
      url={https://arxiv.org/abs/2510.25510}, 
}

@inproceedings{reimers-gurevych-2019-sentence,
    title = "Sentence-{BERT}: Sentence Embeddings using {S}iamese {BERT}-Networks",
    author = "Reimers, Nils  and
      Gurevych, Iryna",
    editor = "Inui, Kentaro  and
      Jiang, Jing  and
      Ng, Vincent  and
      Wan, Xiaojun",
    booktitle = "Proceedings of the 2019 Conference on Empirical Methods in Natural Language Processing and the 9th International Joint Conference on Natural Language Processing (EMNLP-IJCNLP)",
    month = nov,
    year = "2019",
    address = "Hong Kong, China",
    publisher = "Association for Computational Linguistics",
    url = "https://aclanthology.org/D19-1410/",
    doi = "10.18653/v1/D19-1410",
    pages = "3982--3992"
}

@inproceedings{wang-etal-2020-rat,
    title = "{RAT-SQL}: Relation-Aware Schema Encoding and Linking for Text-to-{SQL} Parsers",
    author = "Wang, Bailin  and
      Shin, Richard  and
      Liu, Xiaodong  and
      Polozov, Oleksandr  and
      Richardson, Matthew",
    editor = "Jurafsky, Dan  and
      Chai, Joyce  and
      Schluter, Natalie  and
      Tetreault, Joel",
    booktitle = "Proceedings of the 58th Annual Meeting of the Association for Computational Linguistics",
    month = jul,
    year = "2020",
    address = "Online",
    publisher = "Association for Computational Linguistics",
    url = "https://aclanthology.org/2020.acl-main.677/",
    doi = "10.18653/v1/2020.acl-main.677",
    pages = "7567--7578"
}

@inproceedings{wang-etal-2025-linkalign,
    title = "{L}ink{A}lign: Scalable Schema Linking for Real-World Large-Scale Multi-Database Text-to-{SQL}",
    author = "Wang, Yihan  and
      Liu, Peiyu  and
      Yang, Xin",
    editor = "Christodoulopoulos, Christos  and
      Chakraborty, Tanmoy  and
      Rose, Carolyn  and
      Peng, Violet",
    booktitle = "Proceedings of the 2025 Conference on Empirical Methods in Natural Language Processing",
    month = nov,
    year = "2025",
    address = "Suzhou, China",
    publisher = "Association for Computational Linguistics",
    url = "https://aclanthology.org/2025.emnlp-main.51/",
    doi = "10.18653/v1/2025.emnlp-main.51",
    pages = "977--991",
    ISBN = "979-8-89176-332-6"
}

@article{10.1145/3725331,
    author = {Xie, Xiangjin and Xu, Guangwei and Zhao, Lingyan and Guo, Ruijie},
    title = {OpenSearch-SQL: Enhancing Text-to-SQL with Dynamic Few-shot and Consistency Alignment},
    year = {2025},
    issue_date = {June 2025},
    publisher = {Association for Computing Machinery},
    address = {New York, NY, USA},
    volume = {3},
    number = {3},
    url = {https://doi.org/10.1145/3725331},
    doi = {10.1145/3725331},
    journal = {Proc. ACM Manag. Data},
    month = jun,
    articleno = {194},
    numpages = {24}
}

@misc{yao2025arctictext2sqlr1simplerewardsstrong,
      title={Arctic-Text2SQL-R1: Simple Rewards, Strong Reasoning in Text-to-SQL}, 
      author={Zhewei Yao and Guoheng Sun and Lukasz Borchmann and Zheyu Shen and Minghang Deng and Bohan Zhai and Hao Zhang and Ang Li and Yuxiong He},
      year={2025},
      eprint={2505.20315},
      archivePrefix={arXiv},
      primaryClass={cs.CL},
      url={https://arxiv.org/abs/2505.20315}, 
}

@article{deepseekv3_2_2025,
  title        = {DeepSeek-V3.2: Pushing the Frontier of Open Large Language Models},
  author       = {DeepSeek-AI and others},
  journal      = {arXiv preprint arXiv:2512.02556},
  year         = {2025},
  eprint       = {2512.02556},
  archivePrefix= {arXiv},
  primaryClass = {cs.CL}
}

@article{deepseekmath2024,
  title        = {DeepSeekMath: Pushing the Limits of Mathematical Reasoning in Open Language Models},
  author       = {Shao, Zhihong and Wang, Peiyi and Zhu, Qihao and Xu, Runxin and Song, Junxiao and Bi, Xiao and Zhang, Haowei and Zhang, Mingchuan and Guo, Daya and others},
  journal      = {arXiv preprint arXiv:2402.03300},
  year         = {2024},
  eprint       = {2402.03300},
  archivePrefix= {arXiv},
  primaryClass = {cs.CL}
}

@misc{openai_o3mini,
  title        = {OpenAI o3-mini},
  author       = {{OpenAI}},
  year         = {2025},
  howpublished = {OpenAI},
  note         = {Accessed 2025-12-28}
}

@misc{openai_gpt5,
  title        = {GPT-5},
  author       = {{OpenAI}},
  year         = {2025},
  howpublished = {OpenAI},
  note         = {Accessed 2025-12-28}
}

@article{qwen3_report,
  title        = {Qwen3 Technical Report},
  author       = {Qwen Team and others},
  journal      = {arXiv preprint arXiv:2505.09388},
  year         = {2025},
  eprint       = {2505.09388},
  archivePrefix= {arXiv},
  primaryClass = {cs.CL}
}

@article{chess_2024,
  title        = {CHESS: Contextual Harnessing for Efficient SQL Synthesis},
  author       = {Talaei, Shayan and Pourreza, Mohammadreza and Chang, Yu-Chen and Mirhoseini, Azalia and Saberi, Amin},
  journal      = {arXiv preprint arXiv:2405.16755},
  year         = {2024},
  eprint       = {2405.16755},
  archivePrefix= {arXiv},
  primaryClass = {cs.CL}
}

@article{chase_sql_2024,
  title        = {CHASE-SQL: Multi-Path Reasoning and Preference Optimized Candidate Selection in Text-to-SQL},
  author       = {Pourreza, Mohammadreza and Li, Hailong and Sun, Ruoxi and Chung, Yeounoh and Talaei, Shayan and Kakkar, Gaurav Tarlok and Gan, Yu and Saberi, Amin and Ozcan, Fatma and Arik, Sercan O.},
  journal      = {arXiv preprint arXiv:2410.01943},
  year         = {2024},
  eprint       = {2410.01943},
  archivePrefix= {arXiv},
  primaryClass = {cs.CL}
}

@article{mars_sql_2025,
  title        = {MARS-SQL: A multi-agent reinforcement learning framework for Text-to-SQL},
  author       = {Yang, Haolin and Zhang, Jipeng and He, Zhitao and Fung, Yi R.},
  journal      = {arXiv preprint arXiv:2511.01008},
  year         = {2025},
  eprint       = {2511.01008},
  archivePrefix= {arXiv},
  primaryClass = {cs.CL}
}

@article{qwen25_coder_report,
  title        = {Qwen2.5-Coder Technical Report},
  author       = {Hui, Binyuan and Yang, Jian and Cui, Zeyu and Yang, Jiaxi and Liu, Dayiheng and Zhang, Lei and others},
  journal      = {arXiv preprint arXiv:2409.12186},
  year         = {2024},
  eprint       = {2409.12186},
  archivePrefix= {arXiv},
  primaryClass = {cs.CL},
  url          = {https://arxiv.org/abs/2409.12186}
}

@online{gcp_vertex_gemini3_flash,
  author  = {{Google Cloud}},
  title   = {Gemini 3 Flash | Generative AI on Vertex AI},
  year    = {2025},
  url     = {https://docs.cloud.google.com/vertex-ai/generative-ai/docs/models/gemini/3-flash},
  urldate = {2025-12-29},
  note    = {Model ID: \texttt{gemini-3-Flash}.}
}

@online{gcp_vertex_gemini3_pro,
  author  = {{Google Cloud}},
  title   = {Gemini 3 Pro | Generative AI on Vertex AI},
  year    = {2025},
  url     = {https://docs.cloud.google.com/vertex-ai/generative-ai/docs/models/gemini/3-pro},
  urldate = {2025-12-29},
  note    = {Model ID: \texttt{gemini-3-pro-preview}.}
}

@online{anthropic_claude_opus45_20251101,
  author  = {Anthropic},
  title   = {Introducing Claude Opus 4.5},
  year    = {2025},
  month   = nov,
  day     = {24},
  url     = {https://www.anthropic.com/news/claude-opus-4-5},
  urldate = {2025-12-29},
  note    = {Mentions API model name \texttt{claude-opus-4-5-20251101}.}
}

\end{document}